\documentclass{aa}  

\usepackage{graphicx}
\usepackage{xcolor}
\usepackage{txfonts}
\usepackage{lipsum}
\usepackage{subcaption}         
\usepackage{lscape}             
\usepackage{placeins}           
                                
\begin{document}

   \title{Observing population III stars in the pristine region}


%
%
%

   \author{Sung Kei Li\inst{1, 2, 3}\corrauth{skli@ifca.es}\fnmsep\thanks{Croucher fellow}        
        \and Jose M. Diego \inst{1} \email{jdiego@ifca.es}
        \and Jose M. Palencia 
        \inst{4, 1}
        \and Erik Zackrisson\inst{5}
        \and Rogier Windhorst\inst{6}
        }

   \institute{Instituto de Física de Cantabria (CSIC-UC), Edificio Juan Jordá, Avenida de los Castros, 39005 Santander, Spain \and
   Department of Physics, The University of Hong Kong, Pokfulam Road, Hong Kong  
   \and The Hong Kong Institute for Astronomy and Astrophysics, The University of Hong Kong, Pokfulam Road, Hong Kong, P. R. China \and David A. Dunlap Department of Astronomy and Astrophysics, University of Toronto, 50 St. George Street, Toronto, Ontario M5S 3H4, Canada 
   \and Observational Astrophysics, Department of Physics and Astronomy, Uppsala University, Box 524, SE-751 20 Uppsala, Sweden
    \and School of Earth and Space Exploration, Arizona State University, Tempe, AZ 85287-6004, USA 
   }

   \date{Received \today}

 
  \abstract{The detection of the long-sought Population III (Pop III) stars is, perhaps, one of the most intriguing missing links in modern astronomy. Caustic crossing events (CCE) -- extreme magnification at the critical curves of galaxy clusters of individual stars at cosmological distances -- are postulated to allow for the direct detection of individual Pop III stars. However, in the usual CCE regime, the abundant existence of intracluster stars as microlenses limits the maximum magnification attainable, thus limiting the theoretical detection of luminous stars to relatively low redshifts, significantly smaller than the redshifts where Pop III stars start forming. Here in this paper, we propose to observe Pop III stars as CCEs in the pristine regions of merging clusters. Pristine regions are defined as areas in the lens where the intracluster light, and hence the corresponding stellar microlenses, is exceptionally low. In these regions, the low density of microlenses allows background stars to achieve magnification closer to the theoretical limit (without microlenses). We show that the maximum magnification in the pristine region scales inversely with the surface mass density of stellar microlenses. Adopting Pop III models, regular James-Webb $\sim29\,$mag observations are sufficient to detect individual Pop III stars in pristine regions, offering a unique observational strategy to detect Pop III stars, especially those with low mass and/or extremely hot surface temperature of $\sim 100,000\,$K, with the incoming cluster observations. We roughly estimate a PCCE detection rate of $\sim \mathcal{O}(10^{-3}-10^{-2})$ per El Gordo-like cluster, and a rate of $\sim \mathcal{O}(0.01-0.1)$ per prior detection of Pop III galaxies with star formation rate of $\sim 0.1-1\,M_{\odot}\rm{yr}^{-1}$ as caustic-crossing galaxies in the pristine region. The intermediate regions between high-redshift merging clusters (such as El Gordo) are naturally devoid of microlenses and constitute exceptional targets for the search of Pop III stars as PCCE events.}


   \maketitle
\nolinenumbers

\section{Introduction}

Galaxy clusters with masses of $\sim 10^{15}\,M_{\odot}$ can theoretically magnify individual stars in background lensed galaxies by a factor of $\mu \sim 10^{6}$, with the maximum magnification inversely proportional to the square root of the star radius \citep{Miralda-Escude_1991}. Such high magnification factors can be found at the critical curve (CC) of massive galaxy clusters. However, this extreme magnification is significantly lowered to $\mu \sim 10^{4}$ by the ubiquitous intra-cluster stars acting as microlenses \citep{Venumadhav_2017, Diego_2018, Oguri_2018, Diego_2019}. Despite this drastic reduction in magnification, near the critical curves the magnification can still be sufficient to detect luminous background stars at cosmological distances. These stars are often identified as (micro)caustic crossing events in deep, multi-epoch {\it Hubble} and {\it James-Webb} space telescope (HST and JWST, respectively) observations \citep[e.g.,][]{Kelly_2018, Kelly_2022_Flashlights, Yan_2023, Fudamoto_2025, palencia2026statisticalstudy100magnified}. The microcaustics are produced by the same microlenses that reduce the magnification at the CC. The large magnification at the CC is redistributed around the much smaller CC that form around the microlenses. Due to the microlenses, the rare but extreme magnification at the galaxy cluster CC is traded for the more numerous but more modest magnification factors around the microlenses.

\begin{figure*}
    \centering
    \includegraphics[width=\linewidth]{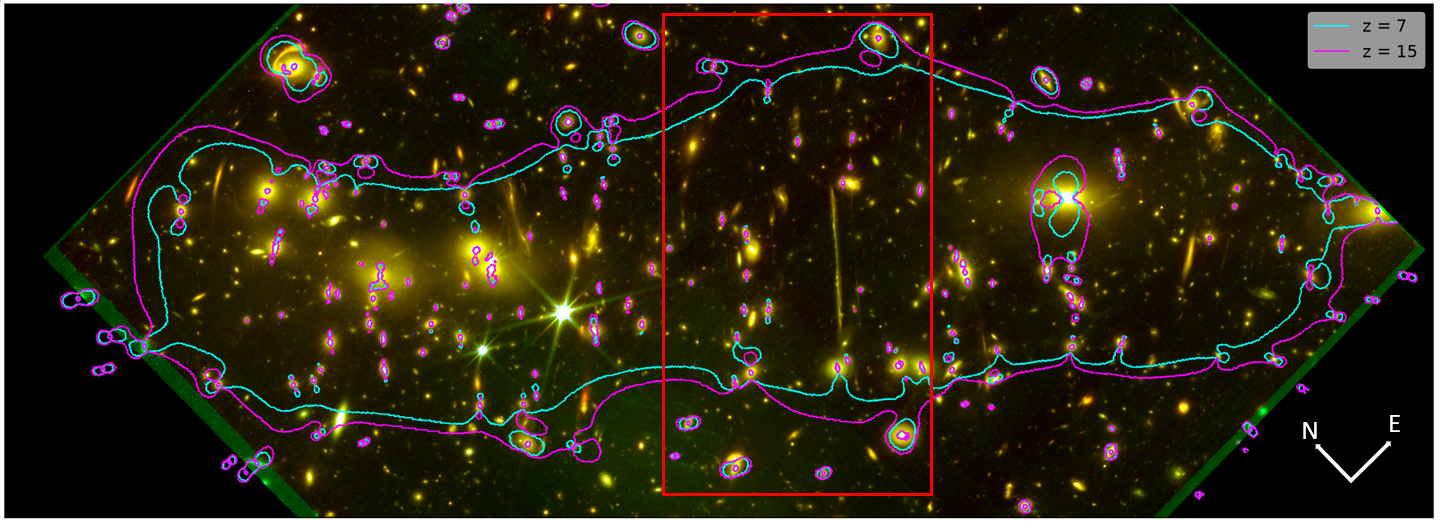}
    \caption{JWST RGB (Red: F356W + F410M + F444W; green: F200W + F277W; blue: F090W + F115W + F150W) image of the cluster ``El Gordo'' ($z = 0.87$), alongside the positions of the cluster critical curves at different source redshifts of $z = 7$ (cyan) and $z = 15$ (magenta) predicted by the model in \citet{Diego_2023_ElGordo}. The pristine region where the lowest density of microlenses intersects the critical curve is highlighted as a red rectangle. }
    \label{fig: Elgordo_pristine}
\end{figure*}

In scenarios studied so far in the literature, the detection of Pop III stars is expected to be very difficult \citep[e.g.,][]{Rydberg_2013, Larkin_2023, Zackrisson_2024, Hassan_2025}. Since CCEs provide a unique opportunity to detect individual stars at cosmological distances, one of the scientific goals of studying CCEs is the direct detection of Pop III stars \citep{Windhorst_2018}. In a usual CCE environment, the abundant microlenses suppress the maximum magnification that is required to detect individual stars at very high redshifts. Here, we propose a unique opportunity to hunt for Pop III stars as CCEs at the ``pristine'' region of the cluster caustic, where the surface mass density of intracluster light is extremely low owing to the merging nature of clusters. In these situations, one cluster amplifies the effect of the other, and a CC can form in a region where each cluster alone could not produce a CC. But the combined effect of the two clusters can produce a CC in a region that is further away from the center of each cluster, and with a smaller contribution from the ICL (and microlenses). If, in addition, the merging cluster is at relatively high redshift, the ICL is still in an early phase of formation and constrained to the innermost region of the cluster, making distant regions relatively empty of microlenses. 
A prime example of a pristine caustic-crossing region would be ACT-CL J0102-4915 \citep[``El Gordo'', $z = 0.87$,][]{Menanteau_2012, Lindner_2014, Diego_2023_ElGordo} as shown in Fig.~\ref{fig: Elgordo_pristine}, where the region between the two merging clusters lacks intracluster light (ICL), leaving behind little to no microlenses while retaining the existence of cluster CC. In these regions, the low abundance of microlenses will increase the magnification attainable for background stars closer to the cluster-only limit of $\sim10^{6}$ without the usual redistribution of magnification from the microlenses. Since Pop III stars are expected only at higher redshift, the CC forms at even larger distances from the center, reducing the abundance of microlenses and increasing the magnification. 

In this paper, we shall show that caustic crossing events at pristine caustic crossing regions, which we dub pristine caustic crossing events (PCCEs) allow for the direct detection of the Pop III star at high redshifts. We shall also briefly discuss observational caveats in the detection of PCCEs. Throughout this work, we adopt the AB magnitude system \citep{Oke_1983}, along with \citet{Planck18} cosmology.

\section{Maximum magnification in the pristine region}
\label{sec: max_mu}

\begin{figure*}
    \centering
    \includegraphics[width=\linewidth]{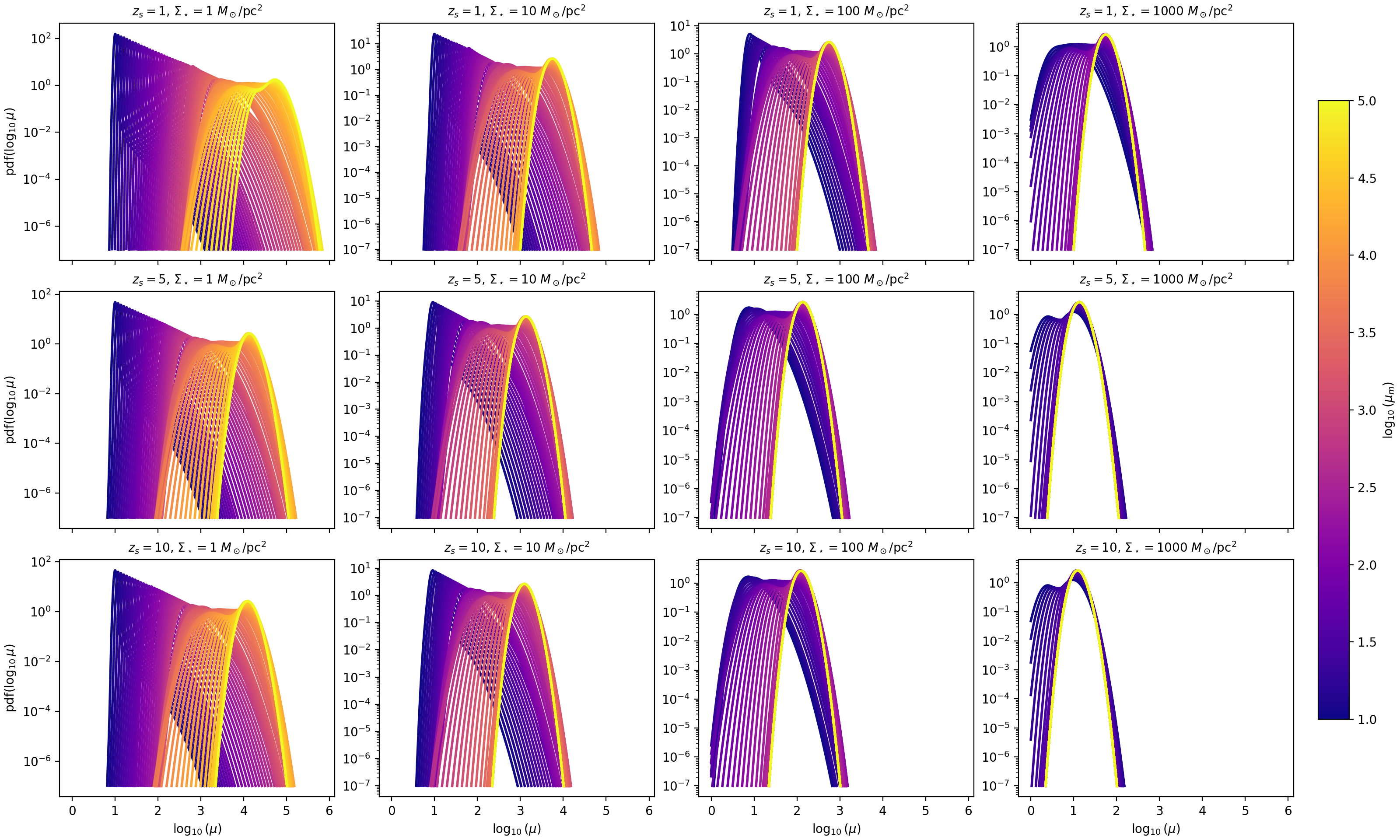}
    \caption{A suite of microlensing PDFs adopting \citet{Palencia_2024}. The three rows from top to bottom show the cases of source redshift at 1, 5, and 10, respectively (corresponding to critical surface mass density of $6715\,M_{\odot}/\textrm{pc}^{2}$, $1633\,M_{\odot}/\textrm{pc}^{2}$ and $1485\,M_{\odot}/\textrm{pc}^{2}$). The four columns, from left to right, shows the cases of microlensing surface mass density equals 1, 10, 100, and 1000$\,M_{\odot}/\textrm{pc}^{2}$. In all these panels, the colour denotes the adopted macromodel magnification from the galaxy cluster, $\mu_{m}$ ($\mu_m$ scales as $1/d$ with $d$ the separation from the CC). One can see that once $\Sigma_{\rm eff} = \Sigma_{\star} \times \mu_{m}$ reaches some threshold, the PDFs converge to a log-normal distribution where they hit the magnification wall, where background sources can no longer gain further magnification even if they are closer to the CC from the cluster. }
    \label{fig: pdfs}
\end{figure*}

At the pristine region of caustic crossing, and for a given surface mass density of microlenses, $\Sigma_*$, when the macromodel magnification, $\mu_m$, is very large and the effective surface mass density of microlenses, $\Sigma_{\rm eff}=\mu_m\times\Sigma_*$, is much greater than the critical surface mass density, $\Sigma_{\rm crit}$, the probability of a background source having some total magnification (macro and micro) follows a log-normal distribution \citep{Diego_2018, Palencia_2024, Diego_2026_Cepheid}. The probability of magnification always converges to this log-normal distribution for sufficiently large values of $\Sigma_{\rm eff}$. As shown by \cite{Palencia_2024}, the peak position of the log-normal converges to a value that is inversely proportional to $\Sigma_*$, and the width tends to a constant value.  We render and show a suite of microlensing probability density functions (PDFs) from \citet{Palencia_2024}\footnote{Even though these simulations where done assuming a pixel size equivalent to a star with radius $R\sim200$ R$_{\odot}$, in the regime of the lognormal PDF for the magnification, the star radius plays a minor role since the net magnification is a collective effect from many microlenses and individual microcaustic crossings, for which the maximum magnification does depend on the star radius, have play a small role in the total magnification} in Fig.~\ref{fig: pdfs} that vary the source redshift, surface mass density of microlenses (from the ICL), $\Sigma_{\star}$, and the macroscopic magnification (from the cluster), $\mu_{m}$ while fixing the radial magnification, $\mu_{r} = 1$. One can see that these PDFs converge to a log-normal distribution when we increase $\mu_{m}$ (or equivalently $\Sigma_{\rm eff}$). Moreover, it is always the PDFs with the lower $\Sigma_{\star}$ that will give rise to the highest possible magnification at any source redshift, even when one increases the macroscopic magnification. With increasing source redshift, the maximum magnification decreases with the critical surface mass density:

\begin{equation}
    \Sigma_{\rm crit} = \frac{c^{2}}{4\pi G}\frac{D_{S}}{D_{L}D_{LS}}, 
\end{equation}

\noindent where $c$ is the speed of light, $G$ the gravitational constant, $D_{S}$, $D_{L}$, and $D_{LS}$ the angular diameter distances to the source, the lens, and between the source and the lens, respectively. 

\begin{figure}
    \centering
    \includegraphics[width=\linewidth]{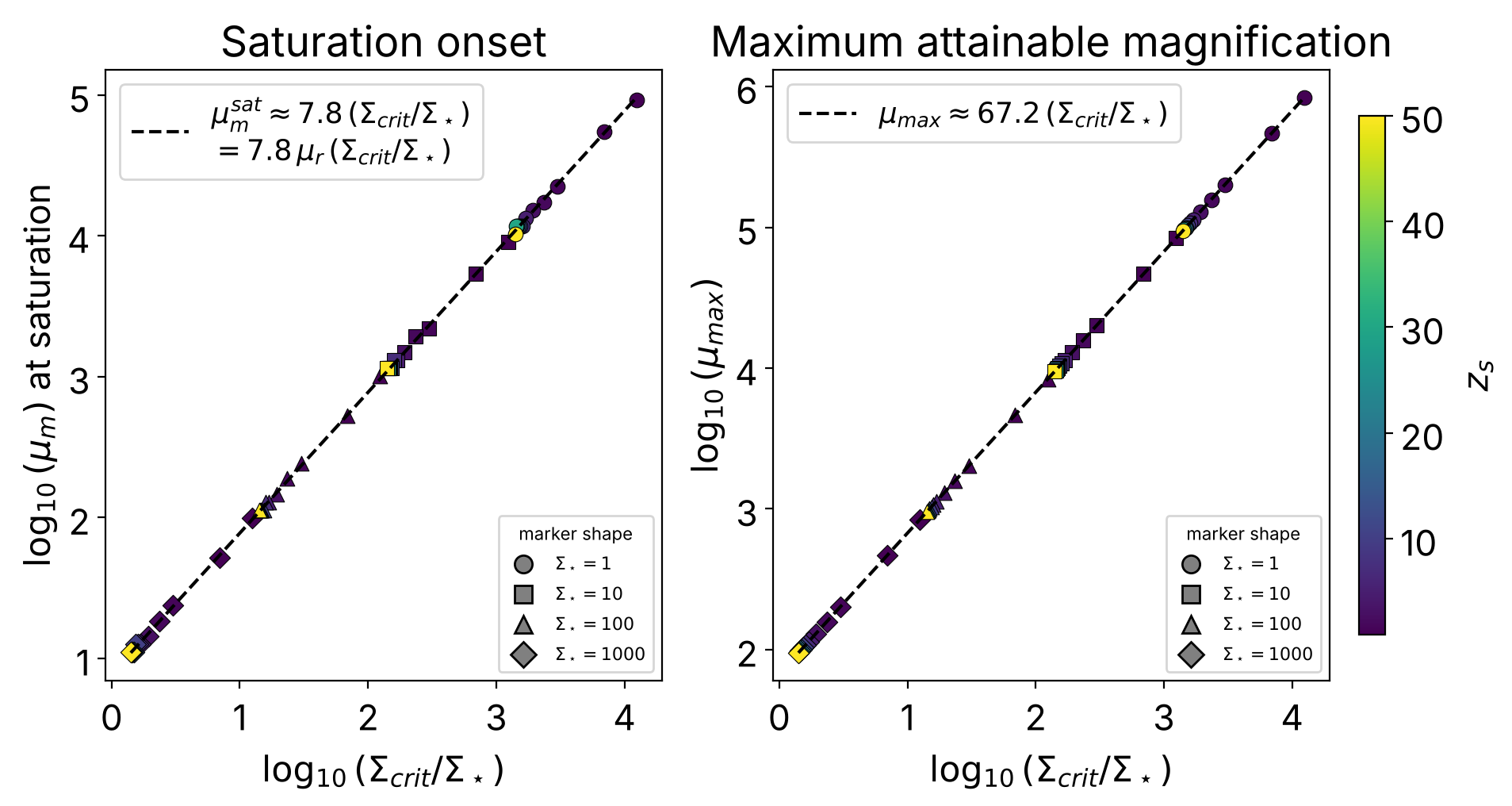}
    \caption{Maximum magnification attainable at the wall of magnification (right) and the corresponding effective surface mass density (left), fitted as a function of $\Sigma_{\rm crit}/\Sigma_{\star}$ with colour denoting the source redshift and marker style denoting the stellar surface mass density adopted. }
    \label{fig: scaling}
\end{figure}

For all the PDFs, we find where the probability drops below $5\sigma$ ($99.99997\%$ for a log-normal distribution) and consider that as the maximum magnification attainable. We also record the effective surface mass density where the PDFs achieve the maximum magnification. The two quantities, at different $\Sigma_{\star}$ and different source redshifts, are plotted in Fig.~\ref{fig: scaling}. Applying a linear fit to the data, one can see that the saturation of magnification, and thus the formation of the ``magnification wall'' \cite{Diego_2026_Cepheid} occurs when the product of $\Sigma_{\star}$ and $\mu_{m}$, known as the effective surface mass density, $\Sigma_{\rm eff}$, is larger than $\sim 7.8$ times the critical surface mass density.

Again applying the fit for the maximum magnification attainable at this log-normal distribution, we have:

\begin{equation}
    \mu_{max} \approx 67 \frac{\Sigma_{\textrm{crit}}}{\Sigma_{\star}},
\label{eqn: mu_max}
\end{equation}

\noindent which is close to the inference in \citet{Diego_2026_Cepheid}. At this magnification ``wall'', the source plane is saturated with microcaustics. Since the area of each microcaustic is orders of magnitude larger than any star, and many microcaustics overlap when $\Sigma_{\rm eff}>>\Sigma_{\rm crit}$, the source star will always be contained within several microcaustics regardless of the star radius. 




\section{Observability and expected rate of Pop III stars as PCCEs}

\begin{figure*}
    \centering
    \includegraphics[width=\linewidth]{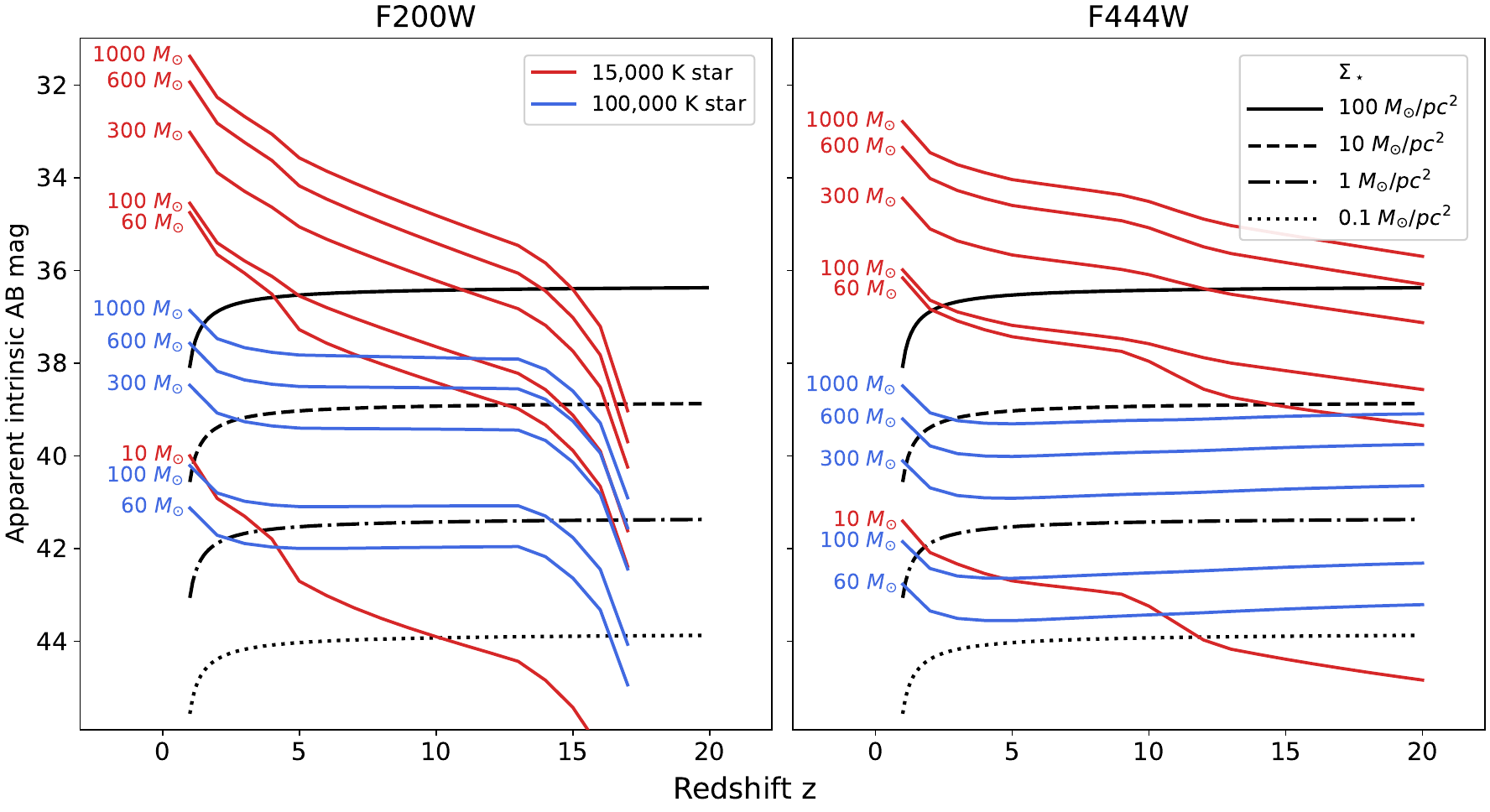}
    \caption{Apparent magnitude of Pop III stars with different initial mass as denoted, and different surface temperature ($15,000\,\rm{K}$ in red, and $100,000\,\rm{K}$ in blue) at different redshifts. Different black lines show the effective detection limit when the nominal $29\,$mag detection limit is boosted by the maximum magnification attainable given different $\Sigma_{\star}$ and redshift (thus $\Sigma_{\rm crit}$), following Eqn.~\ref{eqn: mu_max}. The left panel shows the case for F200W (the NIRCAM wide-band filter with the highest sensitivity), and the right panel shows the case for F444W (longest wavelength NIRCAM wide-band filter). The negative  K correction in the case of the hot star results in a constant sensitivity in the range $4\lesssim z \lesssim 14$. At higher redshifts the Ly-$\alpha$ peak falls in F200W reducing the flux in this filter as the redshift keeps increasing.}
    \label{fig: dimmest_star_detectable}
\end{figure*}

To investigate whether PCCEs allow for the detection of Pop III stars, we first adopt and show the observed apparent magnitudes derived for \citet{Costa_2025} Pop III stars using the Muspelheim models \citep{Zackrisson_2024}  in Fig.~\ref{fig: dimmest_star_detectable} with different surface temperatures ($15,000\,\rm{K}$ in red, and $100,000\,\rm{K}$ in blue) and initial masses (annotated) at F200W (NIRCAM wide-band filter with the highest sensitivity, left panel) and F444W (NIRCAM wide-band filter with the longest wavelength, right panel) respectively. Notice that the curves for F200W vanish after $z\approx 17$, as the Pop III stars can no longer be detected in the passband. Then, we overlay black curves that represent the effective detection limit under a nominal JWST depth of $\sim 29\,$mag, when background sources attain the maximum magnification estimated through Eqn.~\ref{eqn: mu_max} at different redshifts and stellar microlensing surface mass densities denoted by line style (see legend in right panel for the different values of $\Sigma_*$). Coloured curves that fall below the black curves mean that detection of such stars is impossible at the specific $\Sigma_{\star}$. For regions with $\Sigma_{\star}\gtrsim 10\,M_{\odot}/\rm{pc}^{2}$, which is common for lower redshift background galaxies \citep[e.g.,][]{Palencia_2025, Li_2025_IMF}, detection of cooler, massive Pop III stars is possible beyond $z>5$ in F200W and F444W. However, to detect either lower-mass Pop III stars at high redshifts $z\gtrsim15$, or extremely hot Pop III stars at $z\gtrsim 5$ where one would expect Pop III stars \citep{Zier_2025}, one must rely on the pristine region, as they require more magnification to be detected in these filters. 

After we have shown that PCCEs allow for the direct detection of Pop III stars, here we estimate the event rate of Pop III PCCEs in cluster observations. We first estimate the comoving number density of Pop III stars alive, assuming a single-mass Pop III initial mass function at mass $M$ that has a lifetime of $\delta t$, with a star formation rate density, $\dot\rho_{\star}$:

\begin{equation}
    n = \frac{\dot\rho_{\star}\,\delta t}{M}\,V(>\mu_m),
\label{eqn: rate}
\end{equation}

\noindent with $V(>\mu)$ the comoving source plane volume that has magnification larger than $\mu_m$. Characterized earlier in Sec.~\ref{sec: max_mu}, the $\mu_m$ at which the pristine region occurs would be given as $7.7\,\Sigma_{\rm crit}/\Sigma_{\star}$. For $5<z<20$, $\mu_m \sim 10^{4}$. Take El Gordo as an example; the source plane area of which $\mu > 20$ is $\sim 60\,\rm{arcsec}^{2}$ \citep{Diego_2023_ElGordo}. Since the area in the source plane with magnification above a certain value $\mu$ scales as $\mu^{-2}$, the pristine region would have a source area that has $\mu_m \gtrsim 10^{4}$ of $\sim 2.5\times 10^{-4}\,\rm{arcsec}^{2}$, and therefore a comoving volume of $\sim 2\times10^{-3}\,\textrm{c}\,\textrm{Mpc}^{3}$. We adopt a redshift-dependent star formation rate density \citep{Chantavat_2023} that yields an average of $\sim 4\times10^{-4}\,M_{\odot}/\rm{Mpc}^{3}/\textrm{yr}$ at our redshift range. With $M = 100\,M_{\odot}$, $\delta t \approx 3\times 10^{6}\,\rm{yr}$ \citep{Costa_2025}, then $n \approx 12\,\rm{c}\,\rm{Mpc}^{-3}$ and $N = n\times V \approx 0.02$ for the redshift range per El Gordo-like cluster. Notice that this is an optimistic estimation -- not all regions with $\mu_m >10^{4}$ are inside the pristine region, nor is a single-mass stellar initial mass function sensible. Furthermore, we can combine the stellar models and the log-normal microlensing PDFs to estimate, given a detection threshold, what is the expected number of stars that are actually detectable rather than just exist in the region. At $\sim 29\,$mag, the expected detection rate would become $\sim3\times10^{-3}$, such that one out of ten $100\,M_{\odot}$ Pop III stars, or $\sim 3$ in every 1000 stars will be detected; for a more aggressive depth of $30.5\,$mag, the expected rate is $\sim 0.01$, half of the $100\,M_{\odot}$ Pop III stars, or $\sim 1$ in every 100 stars will be detected.

The scenario above is consistent with earlier estimations, making the detection of Pop III stars difficult, but not impossible \citep[e.g.,][]{Windhorst_2018, Zackrisson_2024, perivolotis2026constraintspopiiisky}. A better strategy is to first find the caustic-crossing high-z galaxies hosting Pop III stars, and then monitor these for PCCEs of individual Pop III stars. In this case, the main change in the calculations is the star formation rate in Eqn~\ref{eqn: rate} that transforms from a small volumetric density to a higher rate specific to the caustic-crossing galaxy being considered. This specific rate can be up to two orders of magnitude larger than considered earlier (up to 0.01--0.1 $ {\rm M}_{\odot}$ yr$^{-1}$ per galaxy). This rate can be increased further by an extra one or two orders of magnitude if one considers massive star clusters \citep{Cai_2015}. After correcting for the small volume of these early galaxies, the probability of detecting a Pop III star in a pristine region increases significantly. For a $z = 7$ galaxy of radius $0.2\,\rm{kpc}$, the volume is $\sim3\times10^{-11}\,\rm{Mpc}^{3}$, so a star formation rate with $\mathcal{O}(0.1-1)\,M_\odot\,\rm{yr}^{-1}$ \citep{Cai_2015}, $M=100\,M_\odot$ and $\delta t\approx3\times10^{6}\,\rm{yr}$ gives $\sim3000-30000$ Pop III stars alive in the galaxy. Since only the part of the galaxy that lies at $\mu_m>10^{4}$ contributes in Eq.~\ref{eqn: rate}, we can estimate this by accessing the El Gordo model \citep{Diego_2023_ElGordo} and checking the width of the caustic region with $\mu_m\gtrsim 10^{4}$ through the source plane inverse scaling relation \citep{Blandford_1986}, $\mu_m = C/\sqrt{d}$, of magnification, $\mu_m$, and separation from the fold caustic, $d$. The constant $C$ can be attained through empirical fitting of the model, and it is $\sim 5$ in the \citet{Diego_2023_ElGordo} model. This yields a width of $\sim 2\times 10^{-7}\,$arcsec in the source plane. Assuming that the caustic crosses the center of this source galaxy, the volume within the pristine region is then $\sim 2 \times 10^{-16}\,\rm{Mpc}^{3}$. The expected number of Pop III stars alive within the source galaxy, and within the pristine region, is then $\mathcal{O}(10^{-2})$ to $\mathcal{O}(10^{-1})$, depending on the actual star formation rate.

In reality, (massive) Pop III stars do not necessarily require the extreme magnification in the pristine region in order to be detected, as microlensing on top of cluster strong lensing can reach $\sim 10^{4}$ magnification \citep{Palencia_2024}, which would be sufficient to detect $z\lesssim 10$ high-mass Pop III stars in certain evolutionary states \citep[typically $\gg 100\ M_\odot$ and the low effective temperatures associated with late stages of evolution;][]{Zackrisson_2024}. Such rates were estimated in \citet{Zackrisson_2024} to be on the order of a few events per hundred clusters -- a factor of a few higher than what our optimistic estimation provides, meaning that PCCEs would not be the primary detection channel of Pop III stars. Nevertheless, PCCEs and their extreme magnification would be very relevant for a few specific cases, as we have already briefly discussed earlier in Sec.~\ref{sec: max_mu}:

\begin{enumerate}
    \item If extremely massive Pop III stars at $M>100\,M_{\odot}$ do not exist, then one would have to rely on the extreme magnification of PCCEs to detect the much fainter low-mass Pop III stars.
   \item If, owing to some combination of effects related to rotation, binary evolution, internal convection and mass loss \citep[e.g.,][]{Yoon_2012,Song_2020}, Pop III stars do not evolve to effective temperatures $\lesssim 15000$ K \citep[where they would otherwise peak in $f_\nu$ brightness;][]{Zackrisson_2024,Hassan_2025}, then PCCEs may be the only way to detect individual Pop III stars directly.
    \item Following the previous point, PCCEs are very relevant to catch the extremely hot phase ($\sim 10^{5}\,$K) of Pop III stars to show that super-hot stars exist at high-z.  
    \item If the Pop III $\dot\rho_{\star}$ peaks at $z\gtrsim 15$ and drops very rapidly towards lower redshifts -- then PCCEs would be helpful since moderate-magnification detections may no longer be possible
    \item In the lower magnification regime, while containing a larger source plane area, individual Pop III stars might blend with the surrounding star cluster and make the identification challenging.
\end{enumerate}

On the most conservative side of the estimation, $\mathcal{O}(10^{-2})$ detection probability means one would need $\sim 100$ El Gordo-like clusters to see one event, which might not be realistic given the relatively low number of such clusters known thus far. Thus, real observation strategies would involve observing a mix of clusters at different dynamical states. Also, the aforementioned rate is a rough estimate of how many stars would locate within the pristine region. Some of the PCCE events could be persistent if they are intrinsically luminous enough, making it difficult to distinguish from star clusters. One must therefore rely on lens model(s) to infer the magnification and determine its true nature. Despite their persistence, they could have attained flux variability when they are swapping around the microcaustic in the pristine region. The amplitude of the flux variation, when compared with the low level expected for star clusters \citep[e.g.,][]{Dai_2021, Li_2024}, could possibly allow for distinguishing between PCCE stars and star clusters. For those that are not persistent, the detection rate would then have to be corrected by the rate at which they swap through the microcaustic, depending on both the relative transverse velocity and the angle between the transverse velocity and the local critical curve. For a merger cluster like El Gordo, the relative transverse velocity is expected to be of $\gtrsim 2000\,\rm{km}/\rm{s}$ \citep[e.g.,][]{wilde2026vlastudydisturbedmassive}, with variation from $\lesssim 1700-1900\,\rm{km}/\rm{s}$ to $\sim 2700\,\rm{km}/\rm{s}$ expected cluster from cluster, depending on the exact merging dynamics \citep[see][]{Windhorst_2018}. Dynamically relaxed clusters, which would have fewer pristine caustic-crossing areas, typically have lower transverse velocities of $\sim 1000\,\rm{km}/s$. A slower transverse velocity means the (P)CCE rate is lower, but each event lasts longer than in the case of a faster transverse velocity \citep{Windhorst_2018}. Observationally, more epochs would be required to capture (P)CCEs in merger clusters as they last shorter, and fewer epochs would be required to capture (P)CCEs in relaxed clusters given their longer time spent in caustics.




\section{Conclusion}

In this paper, we show how the largest possible magnification factors ($\sim 10^5$) can be reached in pristine regions, where microlenses minimally perturb the galaxy cluster caustic.  Then we propose the idea of detecting individual Pop III stars in the pristine region of merging clusters, where there are few to no ICLs, as PCCEs. We characterize the maximum magnification background stars can attain in the pristine region as a function of critical surface mass density (thus source redshift) and the abundance of stellar microlenses. We found that the maximum magnification attainable by background stars scales inversely proportional to the abundance of stellar microlenses. This occurs when the effective surface mass density reaches $\sim 7.8$ times the critical surface mass density, where the PDF of magnification converges to a log-normal, and the maximum magnification saturates at the ``magnification wall''. In this regime, increasing the macroscopic magnification no longer increases the probability of having even higher total magnification, and the maximum magnification is set by $\Sigma_*$.

We adopted \citet{Zackrisson_2024} models for \citet{Costa_2025} Pop III stars  to inspect which class of Pop III stars are only detectable in the PCCE regime. While Pop III stars with a low surface temperature of $\sim 15,000\,$K can be detected at a regular CCE surface mass density of $\gtrsim 10\,M_{\odot}/\textrm{pc}^{2}$, lower-mass Pop III stars at higher redshift, and extremely hot (and blue) Pop III with surface temperature $\gtrsim 100,000\,$K can only be detected as PCCE in JWST. We provide a rough estimate of the expected number of PCCE per El Gordo-like cluster based on the comoving volume and star formation rate density at $5< z< 20$, finding that the expected rate is of order $\mathcal{O}(0.001-0.01)$ at an observational depth of $\sim 29-30.5\,$mag, or, $\mathcal{O}(0.01-0.1)$ provided that a caustic-crossing star-forming (star formation rate of $\mathcal{O}(0.1-1)\,M_{\odot}\,\rm{yr}^{-1}$) high-z galaxy that hosts Pop III stars have been identified in advance. Although this rate is lower than the few events per hundred clusters expected for regular CCEs \citep{Zackrisson_2024}, such that PCCE would not be the primary channel in detecting Pop III stars, the unique capability in searching for lower-mass and/or extremely hot Pop III stars would be complementary to regular Pop III CCEs in fully exploring the census of Pop III stars.

El Gordo is not the only cluster with a pristine region -- other merging clusters such as PSZ1 G097.93+19.46 \citep[e.g.,][]{Cerny_2026}, and the outskirts of regular clusters (where images of extremely high redshift lensed sources would locate), contain pristine regions where PCCE detection is possible. Although the PCCE cross section per cluster is very small, {\it Euclid} is expected to provide a pioneering catalog for suitable clusters to hunt for PCCEs with JWST follow-up imaging. This, in combination with ongoing JWST programs such as Strong LensIng and Cluster Evolution (SLICE, GO-5594, PI: Mahler) and Vast Exploration for Nascent, Unexplored Sources (VENUS, GO-6882, PI: Fujimoto), will return tens, if not hundreds, of high-resolution cluster images that are favorable for searches of PCCEs and individual Pop III stars.

\begin{acknowledgements}
S.K.L. acknowledges the generous support of the Croucher Foundation through the Croucher Postdoctoral Fellowship.
J.M.D. acknowledges the support of projects PID2022-138896NB-C51 (MCIU/AEI/MINECO/FEDER, UE) Ministerio de Ciencia, Investigaci\'on y Universidades and SA101P24. E.Z. acknowledges project grant 2022-03804 from the Swedish Research Council (Vetenskapsr\aa{}det).
R.A.W. acknowledges support from NASA JWST Interdisciplinary Scientist grants NAG5-12460, NNX14AN10G and 80NSSC18K0200 from GSFC.
The Instituto de Física de Cantabria (IFCA) has been accredited as Unidad de Excelencia María de Maeztu, a distinction awarded by the Spanish State Research Agency for the period 01/06/2026 - 31/05/2030. This publication is part of grant «Excelencia María de Maeztu» CEX2025-001620-M, funded by MICIU/AEI/10.13039/501100011033. 

We acknowledge the use of the following programs: Astropy \citep{astropy:2013, astropy:2018, astropy:2022}, Numpy \citep{numpy}, and Matplotlib \citep{matplotlib}.

This research is based on observations made with the NASA/ESA {\it James-Webb} Space Telescope obtained from the Space Telescope Science Institute, which is operated by the Association of Universities for Research in Astronomy, Inc., under NASA contract NAS 5-26555. These observations are associated with program GTO-1149. The data are available at MAST: 10.17909/942c-9c60.

\end{acknowledgements}

%
\bibliographystyle{aa} 
\bibliography{bib} 







   
  




\end{document}